\magnification = \magstep1
\vsize = 22.5 true cm
\hsize = 16 true cm
\baselineskip = 24 true pt
\centerline {\bf {PATH INTEGRAL FORMULATION OF THE CONFORMAL}} 
\centerline {\bf {WESS-ZUMINO-WITTEN $\rightarrow$ LIOUVILLE REDUCTION}} 
\bigskip
\bigskip
\centerline {L. O'Raifeartaigh and V. V. Sreedhar} 
\centerline {School of Theoretical Physics}
\centerline {Dublin Institute for Advanced Studies} 
\centerline {10, Burlington Road, Dublin 4}
\centerline {Ireland}
\bigskip
\bigskip
\centerline {\bf {Abstract}}
The quantum Wess-Zumino-Witten $\rightarrow$ Liouville
reduction is formulated using the phase space path  
integral method of  Batalin, Fradkin, and Vilkovisky, adapted to 
theories on compact two dimensional manifolds.  
The importance of the zero modes of the Lagrange multipliers 
in producing the Liouville potential and the WZW anomaly, and 
in proving gauge invariance, is emphasised. A previous problem  
concerning the gauge  
dependence of the Virasoro centre 
is solved.  
\vfill
\hfill DIAS-STP-98-04
\vfil\eject
In the course of the past decade the classical Hamiltonian reduction of 
Wess-Zumino-Witten (WZW) theories to Toda theories using first class 
constraints, and the concomitant reduction of Kac-Moody algebras to W-algebras, 
has been formulated in considerable detail [1]. The quantised version of the 
reduction process has also been considered, but  mainly within the framework of 
canonical quantisation [2]. The elegance of the classical reduction process 
suggests, however, that the most natural framework for quantisation  
is through the path integral. Accordingly, 
in this paper, we present the path integral formulation for the 
quantisation of the simplest WZW $\rightarrow$ Toda reduction, namely the 
reduction of the $SL (2, R)$ WZW theory to the Liouville theory. More general   
cases may be dealt with in an analogous fashion and will be considered 
later. 

The path integral reduction process presents a few  
subtleties that may make it worthwhile to present our results in some  
detail. The main point is that since it is not possible to choose 
configurations such that both the kinetic term and the potential of the 
Liouville Action are simultaneously finite on a non-compact manifold, 
the WZW theory, to be reduced, must be defined on a compact manifold.   
As a consequence of this, the Lagrange multipliers (gauge fields) 
which impose the constraints, have zero modes. It is the integration over   
these zero modes that produces the Liouville potential and the WZW anomaly 
{\it i.e.} the shift $k\rightarrow k-2$ of the WZW coupling constant, in 
the reduced quantum theory. The zero modes are also crucial for proving the 
Fradkin-Vilkovisky theorem regarding the gauge independence of the 
reduction. They are also important for resolving an  earlier ambiguity [3] 
regarding the 
argument ($k~{\hbox {or}}~k - 2$) in the expression for the Virasoro centre 
in two different gauges.  

A second, more technical, aspect of the path integral reduction 
concerns the gauge fixing. Since one of the gauges in which we are 
interested is the WZW gauge in which the Lagrange multipliers are 
set equal to zero (the analogue of the temporal gauge in QED), the 
standard Faddeev-Popov (FP) method does not quite suffice. A more 
general method for quantising constrained systems, namely the 
Batalin-Fradkin-Vilkovisky (BFV) formalism [4], needs to be used. 
In addition to this, since the manifold is required to be compact, 
a curved space generalisation of the standard BFV formalism is desirable. 
The formalism also needs to be modified to take into consideration the 
fact that the constraints are chiral. A suitable refinement of the BFV 
formalism which takes these requirements into account is introduced 
in this paper and it indeed allows the path integral reduction 
to proceed in an elegant and gauge independent manner. 

We begin by summarising the classical WZW $\rightarrow $ Liouville reduction.
The WZW model is defined on a two dimensional compact manifold $\partial\Sigma$ 
by the Action [5]    
$$S = k\int_{\partial\Sigma} Tr~(g^{-1}d g)
\cdot (g^{-1}d g) - {2k\over 3}\int_\Sigma Tr~ 
(g^{-1}d g)\wedge (g^{-1} d g)\wedge (g^{-1}d g)
\eqno(1) $$
Here $g\in G\equiv SL(2, R)$ because this is the WZW group that leads to 
the Liouville theory [1]. 
The two dimensional manifold is parametrised by the light-cone coordinates 
$z_r$ and $z_l$ defined by 
$z_r = {z_0 + z_1 \over 2},~z_l = {z_0 - z_1\over 2}$ respectively. 
The Action is invariant under 
$g\rightarrow gu(z_r ),~ g\rightarrow v(z_l )g $
where $u(z_r), v(z_l) \in G$.  
The conserved Noether currents which generate the above transformations, 
$J_r = -(\partial_r g)g^{-1}, ~J_l = g^{-1}(\partial_l g)$,
take their values in the Lie algebra. 
In order to set up the Hamiltonian formalism, we introduce the Gauss 
decomposition for the group-valued field $g$
$$g = \exp {(\alpha\sigma_+)} \exp {(\beta\sigma_3)} \exp {(\gamma\sigma_-)}
\eqno(2)$$ 
where $\sigma_\pm$ and $\sigma_3$ are the standard generators of the 
$SL(2, R)$ Lie algebra.  
As is well-known, the Gauss decomposition is not valid globally. This issue 
has been dealt with in detail in [6]. For simplicity, we restrict  
our present considerations to the coordinate patch that contains the 
identity. Similar results hold for the other patches.  
In terms of the local coordinates $\alpha , \beta ,\gamma $ on the group 
manifold the Action can be rewritten as 
$$S = k\int d^2z~ \Bigl[{1\over 2} (\partial_r\beta )(\partial_l\beta ) + 
(\partial_l\alpha )(\partial_r\gamma )e^{-\beta}\Bigr] \eqno (3)$$
The momenta canonically conjugate to $\alpha , \beta , \gamma $ respectively 
are  
$$\pi_\alpha  = 
k (\partial_r\gamma )e^{-\beta}~~~ 
\pi_\gamma =   
k (\partial_l\alpha )e^{-\beta }~~~ 
\pi_\beta  = 
k\partial_0\beta \eqno(4)$$ 
The canonical Hamiltonian density $H$ is 
$$ H = {1\over 2k}\pi_\beta^2 +  {k\over 2}(\beta ' )^2 + 
{1\over k}\pi_\alpha\pi_\gamma e^{\beta} + \pi_\alpha
\alpha ' - \pi_\gamma \gamma ' \eqno(5)$$
The currents can be expanded in the basis of the Lie algebra and the 
various components can be read off from the following equations 
$$\eqalign{{\pmatrix {J_r^+ \cr J_r^3 \cr J_r^- }} = 
{\pmatrix { 1 & -2\alpha & -\alpha^2e^{-\beta} \cr 
0 & 1 & \alpha e^{-\beta} \cr 0 & 0 & e^{-\beta}}}{\pmatrix 
{\partial_r\alpha\cr \partial_r\beta\cr \partial_r\gamma} }\cr 
{\pmatrix {J_l^+ \cr J_l^3 \cr J_l^- }} = 
{\pmatrix { e^{-\beta} & 0&0\cr \gamma e^{-\beta}&
1&0\cr 
-\gamma^2 e^{-\beta}& -2\gamma & 1}}
{\pmatrix {\partial_l\alpha \cr
\partial_l\beta \cr \partial_l\gamma }}} \eqno(6)$$ 
The currents may also be expressed completely in terms of the phase space 
variables $\alpha ,\beta ,\gamma$ and their conjugate momenta using the  
relations in Eq.(4).  Further, by using canonical Poisson brackets  
for the phase space variables {\it viz.}
$\{\alpha (z) ,\pi_\alpha (z')\} = \{\beta (z),\pi_\beta (z')\} = 
\{\gamma (z) ,\pi_\gamma (z')\} = \delta (z-z')$, 
the rest being zero, we can check explicitly that the currents satisfy two 
independent copies of the standard $SL(2,R)$ Kac-Moody algebra.  
In terms of the currents, the Hamiltonian density $ H$ can be written in
the Sugawara form 
$$H = {\cal T}_r + {\cal T}_l~~ {\hbox {where}}~~{\cal T}_r = 
{1\over 2}\{J_r^+J_r^- + (J_r^3)^2\}~~{\hbox {and}}~~{\cal T}_l = \{J_l^+J_l^- +
 (J_l^3)^2\} \eqno(7)$$
The constraints we want to impose are 
$$ \Phi_r \equiv J_r^- - m_r = \pi_\alpha - m_r  = 0, ~~~\Phi_l\equiv J_l^+ - 
m_l = \pi_\gamma - m_l = 0 \eqno(8)$$
where $m_r $ and $m_l $ are constants. 
Upon imposing the constraints (8) on the classical Hamiltonian density 
(5) of the $SL(2, R)$ WZW model, we get, apart from boundary terms, 
$$ H_{reduced} = {1\over 2k}\pi_\beta^2 + {k\over 2}(\beta ')^2 + {m_r 
m_l\over k}  e^\beta  \eqno(9)$$ 
This is easily recognised as the expression for the Hamiltonian density of the  
classical Liouville theory. 

As is well-known [1], the constraints in (8) are not consistent with the 
conformal invariance,  
defined by the two Sugawara Virasoro operators  in (7),  
because the currents $J_r^-$ and $J_l^+$ are  
spin one fields.  Hence, the Virasoro generators in (7) are replaced by  
the `improved' generators $T_r = {\cal T}_r - \partial_r J_r^3$ and $T_l
= {\cal T}_l + \partial_l J_l^3$. As will be seen later, this improvement may  
be implemented by coupling the currents to a background metric in a specific, 
non-minimal, way.      
With respect to the conformal group generated by the improved Virasoros, the 
currents 
$J_r^-$ and $J_l^+$ are conformal scalars {\it i.e.} they  have   
conformal weights (0, 0).   
The constraints in (8) are, therefore, compatible with this   
conformal group.  

The currents $J_r^+$ and 
$J_l^-$ now have conformal weights (0, 2) and (2, 0) respectively.  
The phase space variables $\alpha $ and $\gamma $ become primary fields 
of conformal weights (0, 1) and (1, 0) respectively, the field $\beta $
becomes a conformal connection, while $e^{\beta} $ becomes a  primary 
field of weight (1, 1) {\it i.e.} it has the opposite conformal weight 
to the volume element $d^2z$ in the two dimensional space.  

For the quantised version, since the constraints are linear in the momenta,   
it is natural to start with the WZW {\it phase space} path integral, namely,  
$$ I (j) = \int d(\alpha\beta\gamma\pi_\alpha\pi_\beta
\pi_\gamma )~e^{-\int d^2z~[\pi_\alpha\dot\alpha + \pi_\beta\dot\beta +
 \pi_\gamma\dot\gamma -  H + j\beta ]}
\eqno(10)$$  
and to use the BFV formalism for the reduction.  
Here the external source, $j$, is 
attached only to $\beta $ since the other variables will be eliminated 
by the reduction.  

We first give a brief sketch of the BFV formalism. Let $p$ and $q$ be any 
set of canonically conjugate variables, H the canonical Hamiltonian, and  
$$Z = \int d(pq)~ e^{-\int dxdt~[p\dot q - H(p, q)]},\eqno(11)$$ 
the phase space path integral which is to be reduced by a set of 
first class constraints $\Phi (q, p)$. Let $A$ be a set of Lagrange 
multipliers, $B$ their canonically conjugate momenta, and $b, \bar c$ 
and $c, \bar b$ be conjugate ghost pairs. Then define the nilpotent BRST 
charge $\Omega$ and the minimal gauge fixing fermion $\bar \Psi$ by  
$$\Omega = \int dx~ [c\Phi + bB] + \cdots , ~~~ 
~~\bar\Psi = \bar c\chi + \bar b A~~~{\hbox {where}}~~~    
\{\Omega , \Omega \} = 0\eqno(12)$$ 
Here the dots refer to terms which involve higher powers of ghost fields 
 (which actually 
do not occur in the present case) 
and    
$\chi (p, q, A, B) $ is a set of gauge-fixing conditions.  
The BFV procedure then consists of inserting the reduction factor 
$$ F = \int d(ABb\bar b c\bar c )e^{-\int dxdt~[\bar b \dot c - \{\Omega , 
\bar \Psi \}]}\eqno(13)$$ 
into the path integral in (11). 
The Fradkin-Vilkovisky theorem states that the reduced path integral  
is independent of the choice of the gauge fixing fermion $\bar\Psi$. There  
are some exceptions to this theorem, mainly because of the Gribov problem [7].
However, for the example we are considering, the gauge group is abelian, and 
the path integral is shown to be independent of the gauge fixing conditions 
by explicit calculation.  
In the definition of the reduction factor above, it is not necessary 
to include the term $B\dot A + {\dot {\bar c}} b$ in the Action
because such a term can always be generated by letting  
$\chi \rightarrow \chi + \bar c\dot A$. 
The standard non-zero Poisson brackets for the variables  
$ \{q (x), p (x')\} = \{A (x), B(x')\} = 
 \{b (x), \bar c (x')\} = 
 \{c (x), \bar b (x')\} = \delta (x - x') $  
imply that 
$$\{\Omega , \bar\Psi\} = (A\Phi + B\chi) - 
(\bar b b - \bar c [FP]c - \bar c [BFV]b)\eqno(14)$$ 
where the FP and BFV terms are defined by 
$$ \{\Phi (x), \chi (x')\} = [FP]\delta (x - x'), ~~
\{B(x), \chi (x')\} = [BFV]\delta (x - x') = -{\partial\chi\over \partial A}
\delta (x - x')\eqno(15)$$ 
Substituting for $\{\Omega , \bar\Psi\}$ in $F$ and doing the 
$\bar b b$ integrations yields 
$$ F = \int d(AB\bar cc) e^{\int dxdt ~[A\Phi + B\chi + \bar c\{[FP] + 
[BFV]\partial_t \}c ] }\eqno(16)$$
Assuming that $\chi$ is independent of the $B$-fields, as is usually the 
case, we may also integrate over them to get 
$$\eqalign{ F &= \int d(A\bar cc)\delta (\chi )~ 
e^{\int dxdt ~[A\Phi + \bar c\{[FP] + 
[BFV]\partial_t \}c ] }\cr  
 &= \int dA \delta (\chi ) det\Bigl([FP] + [BFV]\partial_t\Bigr)~
 e^{\int dxdt ~[A\Phi ]} } 
 \eqno(17)$$
Note that if $[BFV]$ is equal to zero, we recover the standard Faddeev-Popov 
insertion [8]. On the other hand, as is clear from (15), it is the presence 
of the $[BFV]$ term that allows the gauge fixing function $\chi$ to depend 
on the Lagrange multipliers. 

We now apply the BFV formalism to the WZW $\rightarrow $ Liouville 
reduction.  
The application will  
differ from the standard BFV formalism  
in two respects. First, since we are dealing with independent left 
handed and right handed constraints, it is convenient to replace the standard 
BFV formalism with a light-cone version. The light-cone version of the BFV 
formalism is introduced by replacing the space and time directions by the 
two branches of the light-cone 
parametrised by the light-cone coordinates $z_r$ and $z_l$, using a different 
branch as the time for each of the two constraints.  
However, since we will use the Euclidean version of the theory in the 
path integral, these coordinates 
actually get converted into holomorphic and anti-holomorphic 
coordinates. Thus  
all the fields will be functions of the holomorphic and anti-holomorphic 
variables and functions which depend only on one variable 
will be  holomorphic or anti-holomorphic functions. Second,   
for reasons already explained, we must work on a compact manifold and thus  
we need a curved space generalisation of the BFV formalism.       

Since the left and right-handed constraints are independent, it is easy to 
see that in the light-cone version, the BFV reduction factor $F$ is just 
the product of two factors $F_l$ and $F_r$ where 
$$ F_l = \int dA_l \delta (\chi_l ) det\Bigl([FP]_l + [BFV]_l\partial_r\Bigr)~
 e^{\int dxdt ~[A_l\Phi_l ]}  
 \eqno(18)$$
and similarly for $F_r$. Furthermore, because $\Phi_l =  
\pi_\gamma - m_l$, we see that the argument in the determinant in (18) is
$$[FP]_l + [BFV]_l\partial_r = -
\Bigl({\partial\chi_l\over\partial\gamma} 
+ {\partial\chi_l\over\partial A_l}\partial_r\Bigr)\eqno(19)$$
According to the BFV prescription, the reduction factor (18) is to be 
inserted into the unconstrained WZW path integral (10). Thus, integrating 
over $\pi_\beta$ and regarding $\beta$ as a background field for the 
time being, the reduced path integral is 
$$I = \int d(\pi_\alpha\pi_\gamma\alpha\gamma A_lA_r )\delta (\chi_l )
\delta (\chi_r )det\Bigl\lbrack{ 
\bigl({\partial\chi_l\over\partial\gamma} 
+ {\partial\chi_l\over\partial A_l}\partial_r\bigr)
\bigl({\partial\chi_r\over\partial\alpha} 
+ {\partial\chi_r\over\partial A_r}\partial_l\bigr)\Bigr\rbrack}
~e^{-S_A} \eqno(20a)$$ 
where $S_A$ is given by 
$$\int d^2z~[{k\over 2}(\partial_r\beta )(\partial_l\beta ) 
+ \pi_\alpha\partial_l\alpha + \pi_\gamma\partial_r\gamma -
{1\over k}\pi_\alpha\pi_\gamma e^\beta - A_l(\pi_\gamma - m_l) -
A_r(\pi_\alpha - m_r )] \eqno(20b)$$ 
Integrating over the momenta $\pi_\alpha$ and $ \pi_\gamma$ gives the 
configuration space version of the BFV path integral for the gauged 
WZW model 
$$I = \int d(e^{-\beta}\alpha\gamma A_lA_r )\delta (\chi_l )
\delta (\chi_r )det\Bigl\lbrack{ 
\bigl({\partial\chi_l\over\partial\gamma} 
+ {\partial\chi_l\over\partial A_l}\partial_r\bigr)
\bigl({\partial\chi_r\over\partial\alpha} 
+ {\partial\chi_r\over\partial A_r}\partial_l\bigr)\Bigr\rbrack}
~e^{-S_G} \eqno(21a)$$ 
where $S_G$ stands for the Action of the gauged WZW model and is given by 
$$S_G = \int d^2z~[{k\over 2}(\partial_r\beta )(\partial_l\beta ) 
+ ke^{-\beta}(\partial_l\alpha - 
A_r)(\partial_r\gamma - A_l) 
+ A_lm_l + A_rm_r ] \eqno(21b)$$ 
Equations (21a,b) are the standard BFV results for the reduced path integral  
in Euclidean coordinates. It is obvious that the Action (21b) is invariant 
under the gauge transformations
$$\alpha \rightarrow \alpha + \lambda_r, ~~~~~A_r \rightarrow A_r + 
\partial_l\lambda_r \eqno(22)$$ 
and similarly for $\gamma$ and $A_l$. 

We can now discuss the 
zero modes of the $A$'s. 
This we can do by taking into account the conformal 
spins $\omega (s_l, s_r)$ of the fields 
$$\omega (e^\beta ) = (1, 1),~\omega (\alpha ) = (0, 1),~
\omega (\gamma ) = (1, 0);~~~~~\omega (A_r) = \omega (A_l) = (1, 1)\eqno(23)$$
The weights of $\alpha , \gamma$ and $e^\beta$ were determined following 
(9) and the natural choice of weights for the $A$ fields above follows 
from the gauge transformations (22).  
Consider, for definiteness,  $A_l$, and decompose it into a maximally 
gauge invariant part $A_l^0$ and its orthogonal complement $\hat A_l$ 
which can be gauged away {\it i.e.} let   
$$A_l = A_l^0 + \hat A_l~~{\hbox{where}}~~\hat A_l = \partial_r\lambda_l,~
~{\hbox{and}}~~\int d^2z~ e^{-\beta}A_l^0\hat A_l = 0 \eqno(24)$$
In the above equation the gauge transformation parameter 
$\lambda_l$ has a conformal weight $\omega (\lambda_l ) = (1, 0)$ and 
the factor $e^{-\beta}$ in the integral comes from the requirement  
that the orthogonality condition be defined in a conformally  
invariant manner. 
Since (24) must be true for arbitrary $\lambda_l$, 
it follows from a simple partial integration that 
$$\partial_r \Bigl(e^{-\beta}A_l^0\Bigr) = 0~~ \Rightarrow  
~~A_l^0 = e^\beta f(z_l)\eqno(25)$$
where $f(z_l)$ is an arbitrary holomorphic (or anti-holomorphic) function.
However, since there are no holomorphic (or  
anti-holomorphic) functions on a compact Riemann surface except the constant 
functions [9], we see that $f(z_l)$ must be constant. Similar results hold for 
$A_r$. Thus there is just one zero-mode for each $A$. The Lagrange multiplier 
fields can therefore be written as  
$$A_l = \mu_l e^\beta + \partial_r\lambda_l,~~~
A_r = \mu_r e^\beta + \partial_l\lambda_r \eqno(26)$$
where $\mu_l$ and $\mu_r$ are constants.

As has already been mentioned, it is desirable to have a curved space
generalisation of the path integral in (21) because the manifold is 
compact. 
The background metric $g^{\mu\nu}$ may be used for this purpose.   
An interesting property of the Action (21b) is that if we use conformal  
coordinates $g_{\mu\nu} = e^{\sigma (x)}\eta_{\mu\nu}$,  
the metric does not appear explicitly. Furthermore,  
this continues to be the case when we change 
from the Sugawara Virasoro to the improved one. 
In particular, since the partial derivatives 
act on the sides of the fields that have conformal weight zero, they 
remain ordinary derivatives {\it i.e.} there is no need to modify them with  
the spin connection $\partial\sigma$. The reason for the invariance under the 
change of Virasoro is that  
the change of $\alpha$ and $\gamma$ from scalars to fields of weights (0, 1) 
and (1, 0) respectively is exactly compensated by the change in $e^\beta$ from 
a conformal scalar to a primary field of weight (1, 1). 

As mentioned earlier, the improvement terms in the Virasoro can be incorporated 
explicitly in the presence of a background metric. This is done by  
adding to the 
Lagrangian density, a term of the form 
$\sqrt g g^{\mu\nu} \nabla_\mu J_\nu^3$, 
which in conformal coordinates
reduces to $(\partial_\mu\sigma )J_\mu^3$ 
apart from a total derivative term. However, since  
the field $\beta $ 
is no longer a scalar but a spin-zero connection, the current $J_\mu^3$ 
is no longer a vector but a spin-one connection. To restore the vectorial 
properties of $J_\mu^3$, it is necessary to let $J_\mu^3 \rightarrow 
J_\mu^3 - \partial_\mu\sigma$. In that case, the cross-terms in $ tr (J^3)^2  
+ (\partial_\mu\sigma )J_\mu^3$ exactly cancel leaving a net addition  
to the Lagrangian density of a Polyakov term $-k(\partial\sigma )^2/2$. 
The Polyakov term cannnot be ignored 
because it is this term that produces the known classical centre 
$c = -k$ for the improved Virasoro algebra according to the standard formula 
$\partial S/ \partial\sigma (x) = cR(x)$, where $R(x)$ is the Ricci 
scalar. Thus strictly speaking we should add a Polyakov term $kR$/2 to the 
Action.

We also have to consider 
the effect of the change of Virasoro on the 
measure in (21a). The factor    
$(e^{-\beta}\alpha\gamma )$ in the measure remains a scalar under the  
change of Virasoro. Hence the curved space generalisation of the 
$\alpha \gamma$ integral 
requires only the usual factor $\sqrt g$. On the other hand, since the 
$A$ fields have weights (1, 1), their measure requires a factor
${1\over \sqrt g}$.  

Substituting (26) 
in the gauged WZW path integral (21a, b), and incorporating the above 
mentioned modifications because of the curved space generalisation, we get
$$\eqalign{I = \int d(\sqrt g &e^{-\beta}\alpha\gamma )
d({1\over\sqrt g} \partial_r\lambda_l\partial_l\lambda_r )\cr &\delta (\chi_l )
\delta (\chi_r )
det\Bigl\lbrack{ 
\bigl({\partial\chi_l\over\partial\gamma} 
+ {\partial\chi_l\over\partial \lambda_l}\bigr)
\bigl({\partial\chi_r\over\partial\alpha} 
+ {\partial\chi_r\over\partial\lambda_r}\bigr)\Bigr\rbrack}
~e^{-{\hat S}_G}\times I_0 }\eqno(27a)$$ 
where ${\hat S}_G $ is the Action for the fluctuations and $I_0$ is the path 
integral for the zero modes. Since the cross-terms between 
the $A^0$ and $\hat A$ terms, as well as the $m_r\hat A_r$ and $m_l\hat A_l$ 
terms are pure divergences,  these terms drop out and ${\hat S}_G$ and $I_0$ 
may be written as   
$${\hat S}_G = \int d^2z~[{k\over 2}(\partial_r\beta )(\partial_l\beta ) 
+ k e^{-\beta}\Bigl(\partial_l(\alpha - \lambda_r)\Bigr)
\Bigl(\partial_r (\gamma - \lambda_l)\Bigr) 
- {k\over 2}(\partial\sigma )^2] \eqno(27b)$$ 
and 
$$I_0 = \int d(\mu_r\mu_l)~e^{-\int d^2z~[ke^\beta\mu_r\mu_l - e^\beta  
(\mu_lm_l + \mu_rm_r)] } = e^{{m_rm_l\over k}\int d^2z~e^\beta}\eqno(27c)$$
respectively. Note that the integral over the zero modes $\mu$ has
 produced the Liouville potential term  
${m_rm_l\over k}e^\beta$. The determinant in (27a) may be simplified  
by using ${\partial\chi_r\over\partial\alpha } + {\partial\chi_r\over \partial 
\lambda_r} = 2{\partial\chi_r\over\partial (\alpha + \lambda_r)}$ and a similar 
expression for $\chi_l$ and $\gamma$. The measure in (27a) then reduces to 
$$ 4d(\sqrt g e^{-\beta}\alpha\gamma )
d({1\over\sqrt g} \partial_r\lambda_l\partial_l\lambda_r )
\delta (\alpha + \chi_r )
\delta (\gamma + \chi_l )\eqno(28)$$ 
Eliminating the $\lambda$'s by means of the delta functions and rescaling 
$\alpha$ and $\gamma$ by a factor 2, the path integral becomes 
$$I = \int d(\sqrt g e^{-\beta}\alpha\gamma )
det ({1\over\sqrt g} \partial_r\partial_l)
~e^{-\int d^2z~[{k\over 2}(\partial_r\beta )(\partial_l\beta ) 
- {m_rm_l\over k} e^\beta + ke^{-\beta}(\partial_l\alpha )(\partial_r\gamma ) - 
{k\over 2}(\partial\sigma )^2]}\eqno(29)$$
The $\alpha\gamma$ part of this integral is just the well-known one 
encountered in the computation of the WZW partition function, namely, 
$$\eqalign{I_{\alpha\gamma} = \int d(\sqrt g e^{-\beta}\alpha\gamma ) 
e^{-k\int d^2z~e^{-\beta}(\partial_l\alpha )(\partial_r\gamma )} &= 
\int d(ac)~e^{-\int d^2z~a\Bigl(g^{-{1\over 4}}(D_l^\beta)^T(D_r^\beta)
g^{-{1\over 4}}\Bigr)c}  \cr
&= det \Bigl({1\over \sqrt g}(D_l^\beta)^T(D_r^\beta)\Bigr)^{-1}}\eqno(30)$$ 
where $a = \sqrt k g^{1\over 4}e^{-{\beta\over 2}} \alpha$,   
$c = \sqrt k g^{1\over 4}e^{-{\beta\over 2}} \gamma$,   
$D^\beta = \partial + (\partial\beta )$ and 
$(D^\beta )^T = \partial - (\partial\beta )$. 
Thus the path integral (29) may be expressed as 
$$I = e^{-k\int d^2z~[{1\over 2}(\partial_r\beta )(\partial_l\beta ) 
- {m_rm_l\over k}e^\beta - 
{1\over 2}(\partial\sigma )^2]}
{det {1\over\sqrt g}(\partial_l\partial_r)\over det {1\over \sqrt g}
(D_l^\beta)^T(D_r^\beta)}\eqno(31)$$ 
As is well-known, the ratio of the determinants in (31) may be written in the 
form [10] 
$${det {1\over\sqrt g}(\partial_l\partial_r)\over det {1\over \sqrt g}
(D_l^\beta)^T(D_r^\beta)} = e^{\int d^2z~[(\partial\beta )^2 
- \sqrt gR\beta ]}\eqno(32)$$ 
The expression (32) will be referred to as the WZW anomaly.  
Inserting (32) in (31) 
and reintroducing the $\beta$-integration we have finally the 
reduced configuration space path integral for $\beta$ 
$$I = \int d(g^{1\over 4}\beta )~e^{-\int d^2z 
[{(k - 2 )\over 2}(\partial_r\beta )(\partial_l\beta ) -  
{m_rm_l\over k}e^\beta + \sqrt g (j +  R)\beta - {k\over 2}(\partial\sigma )^2]}
\eqno(33)$$
This is just the Liouville path integral in a curved background. Writing 
$\beta = \phi - \sigma $, where $\phi$ is a scalar field, (33) becomes 
$$I = \int d(g^{1\over 4}\phi )~e^{-\int d^2z 
\bigl[{(k - 2 )\over 2}(\partial_r\phi )(\partial_l\phi ) 
- {m_rm_l\over k}\sqrt g e^\phi + \sqrt g \bigl( 1 + (k - 2)\bigr) R\phi 
+ \sqrt g j\phi\bigr]}
\eqno(34)$$
where the terms that depend purely on $\sigma$, including the Polyakov term,
 have cancelled (except for 
the $j\sigma$ term which we have dropped). It is well-known that the 
Virasoro centre for this theory has the form 
$$c(k - 2 ) = {\hbar\over 6} + \hbar(k - 2)\Bigl[\bigl( 1 +
 {1\over (k - 2 )}\bigr)\Bigr]^2 =   
{\hbar\over 6} + (\kappa - 2\hbar )\Bigl[\bigl( 1 +
 {\hbar\over (\kappa - 2\hbar )}\bigr)\Bigr]^2\eqno(35)$$  
where, the $\hbar$/6 comes from the Weyl anomaly for one scalar field, 
and, to separate the quantum effects, we have recalled from (1) that 
$k = \kappa /\hbar$ where $\kappa \sim 1$. 
The results are independent of the choice of the gauge 
fixing conditions -- as predicted by the Fradkin-Vilkovisky theorem. 

Note that the WZW anomaly (ratio of determinants) was produced by the 
fact that the integration over the $\hat A$ fields is restricted by the 
condition that $\hat A$ be a gradient. This restriction is non-trivial 
because the gradients do not form a complete basis on account of the 
existence of the zero modes. Had $A$ been free, we could have integrated 
directly over $A$ in (21) and there would have been no WZW anomaly (although 
there would still be a Weyl anomaly). Thus the WZW  
anomaly, like the Liouville potential, originates in the zero modes. 
The presence of the WZW anomaly means that although the classical reduction 
converts the WZW theory into a Liouville theory with coupling constant $\kappa$,
the quantum reduction converts it into a Liouville theory with coupling 
constant $\kappa - 2\hbar$.

We would now like to investigate what happens when we neglect the zero modes, 
as was done in previous investigations [12], where actually only two extremal 
gauges, namely the Liouville gauge $\chi_r = \alpha , \chi_l = \gamma$ 
and the WZW gauge $\chi_r = A_r , \chi_l = A_l$ were used. The key 
equation for comparison is (21), just prior to the separation of the $A$ fields
into their zero mode and gauge variant parts. In the Liouville gauge, one sees 
from (15) that $[FP] = -1$ and $[BFV] = 0$. Hence the $\alpha$ and $\gamma$ 
fields are eliminated by the delta functions and the integration over the 
$A$ fields produces the Liouville potential. However, if the $\hat A$ fields 
are not restricted to be gradients, there is no WZW anomaly and one obtains 
the Virasoro centre $c(k)$ instead of $c(k - 2)$ in (35).  
In the WZW gauge, $A = 0$, one sees from (15) that $[FP] = 0$ and $[BFV] = -1$,
which, incidentally, shows the necessity of using the BFV formalism for 
considering this gauge. This time the $A$ fields are eliminated by 
the gauge fixing delta functions and we are left with a direct product 
of the WZW path integral and the path integral for the ghosts. The Liouville 
potential is not produced in this case. Nevertheless, the Virasoro centre 
which is just the sum of the WZW centre, the ghost centre and the classical 
improvement centre, is equal to the correct (gauge independent) value $c(k - 
2)$ [3].  

Thus we have the remarkable result that, if the zero modes are neglected, 
the Liouville gauge produces the correct potential and the wrong centre, 
while the WZW gauge produces the correct centre and the wrong potential. 
This disagreement between the results in the two gauges is contrary to 
what one might expect from the Fradkin-Vilkovisky theorem. Our analysis 
shows that if the zero modes are taken into account we obtain the Liouville 
potential and the same centre  
in all gauges. This illustrates the  
importance of the zero modes for the validity of the 
Fradkin-Vilkovisky theorem.  
\bigskip
We would like thank L. Feh\'er, M. Fry, I. Sachs, and 
I. Tsutsui for interesting discussions. 
\vfil\eject
\centerline {\bf {REFERENCES}}
\bigskip
\item {1. } F. A. Bais, T. Tjin and P. Van Driel, Nuc. Phys. {B 357} (1991) 
632; V. A. Fateev and S. L. Lukyanov, Int. J. Mod. Phys. {A3} (1988) 507;
S. L. Lukyanov, Funct. Anal. Appl. {22} (1989) 255; P. Forg\'acs, A. Wipf,
J. Balog, L. Feh\'er, and L. O'Raifeartaigh, Phys. Lett. {B227} (1989)214;
J. Balog, L. Feh\'er, L. O'Raifeartaigh, P. Forg\'acs and A. Wipf, Ann. Phys. 
{203} (1990) 76; Phys. Lett {B244}(1990)435; L. Feh\'er, 
L. O'Raifaertaigh, P. Ruelle, I. Tsutsui and A. Wipf, Phys. Rep. { 222}
No. 1, (1992)1; P. Bouwknegt and K. Schoutens Eds. W Symmetry (Advanced 
series in mathematical physics, 22) World Scientific, Singapore (1995).  

\item {2. } A. Bilal and J. L. Gervais, Phys. Lett {B206} (1988) 412;
Nuc. Phys. { B314} (1989) 646; { B318} (1989) 579; O. Babelon, Phys. 
Lett. { B215} (1988) 523, T. Hollowod and P. Mansfield, Nuc. Phys. 
{ B330} (1990) 720; P. Mansfield and B. Spence, Nuc. Phys. { B362}  
(1991) 294; P. Bowcock and G. M. T. Watts, Nuc. Phys. { B379} (1992) 63;
C. Ford and L. O'Raifeartaigh, Nuc. Phys. { B460} (1996) 203. 

\item {3. } L. O'Raifeartaigh in 1992 CAP/NSERC Summer Institute in 
Theoretical Physics, Quantum Groups, Integrable Models and Statistical 
Systems Eds. J. LeTourneux and Luc Vinet (World Scientific 1993). 

\item {4. } I. A. Batalin and E. S. Fradkin, in: Group Theoretical Methods 
in Physics, Vol II (Moscow, 1980); I. A. Batalin and G. Vilkovisky, 
Phys. Lett. { B69} (1977) 309; E. S. Fradkin and G. A. Vilkovisky, 
Phys. Lett. { B55} (1975) 224; M. Henneaux, Phys. Rep { 126} No.1, 
(1985) 1. 

\item {5. } E. Witten, Comm. Math. Phys. { 92} (1984) 483; P. Goddard and 
D. Olive, Int. J. Mod. Phys. { A1} (1986) 303; P. Di Francesco, P. Mathieu,
and D. S\'en\'echal, Conformal Field Theory, Graduate Texts in Contemporary 
Physics, (Springer-Verlag New York, Inc. 1997).     

\item {6. } I. Tsutsui and L. Feh\'er, Prog. Theor. Phys. Suppl. { 118} 
(1995), 173. 

\item {7. } J. Govaerts, Int. J. Mod. Phys. A4 (1989) 173, Int. J. Mod. Phys. 
A4 (1989) 4487, Class. Quant. Grav. 8 (1991) 4487.  

\item {8. } L. D. Faddeev, Theor. Math. Phys. { 1} (1970) 1;
 L. D.  Faddeev and  A. A. Slavnov, Gauge Fields, introduction 
to quantum theory. (Benjamin Cummins, Reading, Massachusetts, 1980).  

\item {9. } Phillip A. Griffiths, Introduction to Algebraic Curves, 
Vol. { 76} Translations of Mathematical Monographs, American 
Mathematical Society (1989). 

\item {10. } M. L\"uscher, Ann. Phys. 142 (1982) 359; H. Leutwyler and 
S. Mallik, Z. Phys. C33 (1986) 205.  

\item {11. } T. L. Curtright and C. B. Thorn, Phys. Rev. Lett. { 48} 
(1982) 1309, E. D' Hoker and R. Jackiw, Phys. Rev. { D26} (1982) 3517; 
Phys. Rev. Lett. { 50} (1983) 1719.  

\item {12. } L. O'Raifeartaigh, P. Ruelle and I. Tsutsui, Phys. Lett. 
{ B258} (1991) 359. 
\vfil\eject\end